\input{aipcheck}

\documentclass[
    ,final            
  ]
  {aipproc}

\layoutstyle{6x9}

\def\a{\alpha} \def\b{\beta}  \def\d{\delta}
\def\e{\epsilon}  \def\g{\gamma}
   
 \def\m{\mu} \def\n{\nu} 
  \def\r{\rho} \def\s{\sigma}

\def\L{\Lambda}

\def\itrema{$\ddot{\scriptstyle 1}$}

 \def\Bar#1{\overline{#1}}

\def\frac#1#2{{\textstyle{#1\over\vphantom2\smash{\raise -.20ex
    \hbox{$\scriptstyle{#2}$}}}}} 

\def\low#1{{\raise -3pt\hbox{${\hskip 1.0pt}\!_{#1}$}}}
\def\du#1#2{_{#1}{}^{#2}} 
\def\-{{\hskip 1.5pt}\hbox{-}}
\def\scst{\scriptstyle}

\def\eqdot{~{\buildrel{\hbox{\LARGE .}} \over =}~}
\def\eqdot{~{\buildrel{\hbox{.}} \over =}~}

\def\eqques{~{\buildrel ? \over =}~}
\def\Hat#1{\widehat{#1}}                        
\def\Bar#1{\overline{#1}}                       

\def\calR{{\cal R}}

\def\doit#1#2{\ifcase#1\or#2\fi}
\def\be{\begin{equation}} \def\ee{\end{equation}}
\def\ba{\begin{array}} \def\ea{\end{array}}
\def\bea{\begin{eqnarray}} \def\eea{\end{eqnarray}}

\def\ap#1#2#3{Ann.~of Phys.~{\bf {#1}} (19{#2}) #3}

\def\nc#1#2#3{Nuovo Cim.~{\bf {#1}} (19{#2}) #3}
\def\ibid#1#2#3{{\it ibid.}~{\bf {#1}} (19{#2}) #3}

\def\prn#1#2#3{Phys.~Rev.~{\bf D{#1}} (20{#2}) #3}
\def\cqgn#1#2#3{Class.~\& Quant.~Gravity {\bf {#1}} (20{#2}) #3}

\newskip\humongous \humongous=0pt plus 1000pt minus 1000pt

\newif\ifdtup

\begin{document}

\title{Nilpotent Spinor Symmetry with \\
Interacting Spin 3/2 Field\footnote{Talk delivered at SUSY06, 
the 14th International Conference on Supersymmetry and the
Unification of Fundamental Interactions, Irvine, CA, June 2006.  
Preprint \#: CSULB-PA-06-3}}

\classification{11.10.Kk, 11.15.Bt, 11.30.Pb, 11.10.Ef}
\keywords{Spin 3/2 Field, Consistent Interactions,
Nilpotent Spinor Charge, Non-Abelian Tensors.} 

\author{Hitoshi Nishino{\hskip 0.01in}%
~~}{}

\author{~Subhash Rajpoot{\hskip 0.01in}}
{address={Dept.~of Physics \& Astronomy,
  California State University, \\
  1250 Bellflower Blvd., Long Beach, CA 90840} }


\begin{abstract}
\medskip

Consistent interactions of spin 3/2 field that realize a nilpotent
spinorial symmetry are presented. Based on our previous results on
purely bosonic non-Abelian tensor with consistent interactions, we
present a new system for interacting spin 3/2 field that realizes
the nilpotent fermionic symmetry.

\end{abstract}

\maketitle


\section{Introduction}



We have recently constructed 
a system of interacting non-Abelian tensor for purely bosonic 
consistent interactions \cite{NRnatensor}.  
Encouraged by this result, we generalized the basic structure
to a fermionic system, in particular with a spin 3/2 field
\cite{NRfermi}, which is different from the well-known supergravity
case. We found that there are indeed consistent interactions of spin
3/2 field, realizing the nilpotent spinor algebra $~\{ Q\du\a I,
Q\du\b J \}=0$.


\section{(I) Non-Abelian Tensor (Bosonic Case)}

The problem with constructing a system with non-Abelian tensor with
consistent interactions can be described  as follows: Consider the
tensor $~B\du{\m\n}I$~ with the adjoint \hbox{index $~{\scst I}$.}
The problem is that its na\itrema ve field strength $~G\du{\m\n\r}I
\equiv 3 D_{[ \m} B\du{\n\r]} I $~ is {\it not} invariant under the
tensorial gauge transformation $~\d_\L B_{\m\n} = 2 D_{[ \m} \L\du
{\n]} I$, because $~\d_\L G\du{\m\n\r} I = + 3 f^{I JK} F\du{[ \m\n}
J \L\du{\r]} K \neq 0$.

A clue to solving this problem lies in the dimensional reduction
technique of Scherk and Schwarz \cite{Scherk}. In the generalized
dimensional reduction in \cite{Scherk}, the original coordinates
$~(\Hat x^{\hat\m})$~ in $~D+E$~ space-time dimensions~ are reduced
into $~(x^\m, y^\a)$, where $~{\scst \m, ~\n,~\cdots
~=~0,~1,~\cdots,~ D-1}$~ for the $~D\-$dimensions, and $~{\scst \a,~
\b,~\cdots~=~1,~2,~\cdots,~E}$~ for the compact $~E\-$dimensional
extra-coordinates.  For example, there arises the field strength
component \bea G_{\m\n\r\a}= 3  D_{[  \m} B_{\n\r] \a} + 6 F\du{[
\m\n} \b B_{\r ] \a\b} \eea
as the field strength of the tensor $~B_{\m\n\a}$~
with both the $~D\-$ and $~E\-$dimensional coordinates.
The important point is the involvement of the Chern-Simons terms
at the end, different from the above-mentioned na\itrema vely-constructed
field strength.

In \cite{NRnatensor}, we mimicked these structures for dimensional
reductions in \cite{Scherk} to solve the non-Abelian tensor problem.
We introduce the set of fields $~(B\du{\m\n} I, C\du\m{I J}, K^{I J
K} ; ~A\du\m I)$~ and their field strengths: $~F_{\m\n} \equiv 2
\partial_{[ \m} A\du{\n ]} I + f^{I J K} A\du\m J A\du\n K$, \bea
G\du{\m\n\r}I & \equiv &  + 3 D_{[ \m}B\du{\n\r]} I
     -3 F\du{[ \m\n} J C\du{\r]} {J I} ~~, \cr
H\du{\m\n}{I J} &\equiv &  + 2 D_{[ \m} C\du{\n]}{I J}
      +F\du{\m\n} K K^{I J K} + f^{I J K}B\du{\m\n} K~~, \cr
L\du\m{I J K} & \equiv & D_\m K^{I J K} - 3 f^{[  I J | L} C\du\m{L |K]} ~~.
\eea
Note that the total number of space-time indices and adjoint indices
in the field strengths $~G,~H$~ and $~L$~ is always four just as in \cite{Scherk}.
These field strengths are now invariant under
the tensorial $~\L\-$gauge transformations
\bea
\d_\L B\du{\m\n} I & = & 2 D_{[ \m} \L\du{\n]} I
    - F\du{\m\n} J \L^{I J}~~, ~~~~
    \d_\L C\du\m{I J} = D_\m \L^{I J} - f^{I J K} \L\du\m K ~~, \cr
\d_\L K^{I J K} & = & 3 f^{ [  I J | L} \L^{L | K ] } ~~, ~~~~
     \d_\L A\du\m I = 0 ~~.
\eea
In other words, we  succeeded in defining invariant field strengths
that are invariant under the $~\L\-$ tensorial transformations.


\section{(II) Fermionic Nilpotent Symmetry}
\smallskip

 The algebra realizing the nilpotent fermionic symmetry  is
dictated by the three (anti)commutators: \bea \{ Q\du\a I , Q\du\b J
\} & = & 0 ~~, \cr [ T^I, Q\du\a J ] & = &  + f^{I J K} Q\du\a K ~~,
\cr [ T^I, T^J ] & = &  + f^{I J K} T^K ~~, \eea
where $~{\scst I, ~J, ~\cdots ~=~ 1, ~2, ~\cdots , ~g}$~ are for
the adjoint representation of an arbitrary gauge group $~G$,
while $~{\scst \a,~\b,~\cdots}$~ are for the Majorana spinors
components in an arbitrary space-time dimensions $~D$.  As the
first anti-commutator in (4) shows, our spinor charges
are nilpotent.  The last equation in (4) is nothing but the
usual non-Abelian generator commutations, while the second one
implies that our spinor charges $~Q\du\a I$~ are in the adjoint
representations.

The required set of fields  can be fixed by mimicking the previous
bosonic non-Abelian tensors.  The correspondence between the bosonic
and fermionic systems is \bea \hbox{Bosonic Case} ~~ &
\longleftrightarrow & ~~
      \hbox{Fermionic Case}  \cr
& ~~~ & \cr
      (B\du{\m\n} I , C\du\m{I J} , K^{I J K} ; ~A\du\m I)
~~ & \longleftrightarrow &~~ ( \psi\du{\m\a} I , \chi\du\a{I J} ; ~A\du\m I)
\eea
Note that the total number of the space-time bosonic, fermionic
and adjoint indices is always three in the present fermionic system,
similarly to the previous bosonic case.
The correspondence between potentials and field strengths in the
present fermionic system is tabulated as
\bea
\hbox{Potentials} & ~~\longleftrightarrow ~~& \hbox{Field Strengths} \cr
& ~~~ & \cr
\psi\du\m{\a I} &~~ \longleftrightarrow~~&  \calR\du{\m\n} {\a I} \equiv
      2 D_{ [  \m} \psi\du{\n ] } {\a\, I} + F\du{\m\n} J \chi^{\a\, I J} \cr
\chi\du\a{I J} & ~~\longleftrightarrow~~&  L\du{\m\a} {I J}
\equiv D_\m \chi\du\a{I J}
      + f^{I J K} \psi\du{\m\a} K \cr
A\du\m I & ~~\longleftrightarrow~~&  F_{\m\n}
      \equiv 2 \partial_{ [  \m} A\du{\n ] } I + f^{I J K} A\du\m J A\du \n K
\eea
All the terms accompanying the leading gradient terms are interpreted
as generalized Chern-Simons terms.

Our fermionic symmetry transformation rule is now
\bea
\d_Q \psi\du\m I & = &  D_\m \e^I ~~, ~~~~ \d_Q A\du\m I = 0 ~~, \cr
\d_Q \chi^{I J} & =  & - f^{I J K} \e^K = - Q^{I J, K L} f^{K L M} \e^M ~~.  
\eea
We can confirm that the commutator $~[ \d_Q(\e_1) , \d_Q(\e_2) ]$~
vanishes, consistently with the anti-commutator in (4).
Amazingly and amusingly, the
field strengths $~\calR$~ and $~L$~ are both invariant under
the fermionic symmetry:
\bea
& \d_Q \calR\du{\m\n} I = 0 ~~, ~~~~
      \d_Q L\du\m {KI J} = 0 ~~.
\eea

The usual gauge transformation is $~\d_T A\du\m I = D_\m \a^I,~~
\d_T \psi\du\m I = - f^{I J K} \a^J \psi\du\m K, ~~
\d_T \chi^{I J} = - 2 f^{[ I | K L} \a^K \chi^{L | J]}$,
under which the field strengths $~\calR$~ and $~L$~ are both
{\it covariant}, as desired: $~\d_T \calR\du{\m\n} I = - 2 f^{I J K} \a^J \calR\du{\m\n} K, ~~\d_T L\du\m{I J} = - 2 f^{[ I | K L} \a^K L\du\m{L | J ]}$.
Needless to say, these transformation rules are consistent with
the algebra (4).

Now that we have established the invariant field strengths $~\calR$~ and $~L$,
it is straightforward to construct invariant actions under both $~\d_Q$~ and usual
gauge transformations $~\d_T$.  Our classical total action $~I_{\rm T}
\equiv I_1 + I_2 +I_3$~ consists of the typical invariant actions
\bea
I_1  & \equiv &  \int d^D x \,  \Big[ + \frac 14 a_0^{-1} f^{I J K}
     (\Bar L\du\m{I J} \g^{\m\n\r} \calR\du{\n\r} K ) \,  \Big]  \cr
  & = &   \int d^D x \,
        \Big[ + \frac 14 ( \Bar\psi\du\m I \g^{\m\n\r} \calR\du{\n\r} I)
        -  \frac 1 4 a_0^{-1} f^{I J K} (\Bar\chi{\,}{}^{I J} \g^{\m\n\r}  L\du\m{K L})
        F\du{\n\r} L \, \Big] { ~~, ~~~~~ ~~~~~} \cr
I_2  & \equiv &  \int d^D x \,  \Big[ + \frac 12 P^{I J, K L} (\Bar\chi{\,}^{I J}
          \g^\m L\du\m{K L} ) \,  \Big]
          = \int d^D x \,  \Big[ + \frac 12 P^{I J, K L} (\Bar\chi{\,}^{I J}
          \g^\m D_\m\chi^{K L} ) \,  \Big]  {~, ~~~~~ ~~~~~}  \cr
I_3  &  \equiv & \int d^D x \,
     \Big[ - \frac 14 (F\du{\m\n}I )^2 \, \Big]  {~~~~~ ~~~~~  }
       (f^{I K L} f^{ K L J} = a_0 \d^{I J}) ~~.
\eea

As noted in  the bosonic case, the invariance of the field strengths
is closely related to the consistency of all the field equations.
This can be seen from the $~\psi_\m$~ and $~\chi\-$field
equations\footnote{The symbol $~\eqdot$~ is for a field equation.
Relevantly, the symbol $~\eqques$~ is for an equation under
question.} \bea \frac{\d I_{\rm T}}{\d\Bar\psi\du\m I} & = &  +
\frac 12 \g^{\m\r\s} \calR\du{\r\s} I
         - \frac 14 \g^{\m\r\s} \chi^{I J} F\du{\r\s} J
              + \frac 14 Q^{I J, K L} \g^{\m\r\s} \chi^{K L}
              F\du{\r\s} J \eqdot 0 ~~,  \cr
\frac{\d I_{\rm T}}{\d\Bar\chi{\,}^{I J}}
& = & + P^{I J, K L} \g^\m L\du\m {K L}
          - \frac 14 a_0^{-1} f^{I J K} \g^{\m\r\s} L\du\m{K L} F\du{\r\s} L \cr
& & + \frac 1 4 a_0^{-1} f^{[ I | K L} \g^{\m\r\s} L\du\m{K L} F\du{\r\s} {|J]}
      \eqdot 0 {~~.  ~~~~~ ~~~~~}
\eea
The question now is whether the $~\psi_\m\-$field equation in (10)
satisfies the consistency equation $~D_\m \big( \d I_{\rm T} / \d\psi\du\m I  \big)  \eqques 0 $~.
Fortunately, our system has a good answer to this question, thanks
to the identity:
\bea
& D_\m \left( \frac{\d I_{\rm T}}{\d\psi\du\m I} \right)
     + f^{I J K} \left( \frac{\d I_{\rm T}}{\d\chi^{J K}} \right) \equiv 0 ~~.
\eea
Up to this stage, no field equation has been used.
Note also that this identity is nothing but
the invariance of the action $~\d_Q I_{\rm T}=0$.  Now,
the first term in (11) vanishes, as the necessary
condition of the $~\chi\-$field equation.
This feature is also associated with the $~F\chi\-$term and $~\psi\-$linear
term in the $~\calR$~ and $~L\-$field strengths.

In passing we note that we have also performed the quantization of
the system \cite{NRfermi} by confirming the BRST invariance.


\section{(III) Conclusions}

We have established consistent interactions for spin 3/2 (vector spinor)
field $~\psi\du\m{\a I}$~ carrying spinorial and adjoint indices $~{\scst \a}$~ and
$~{\scst I}$.  The main breakthrough is the discovery of the right definition of the field strength $~\calR\du{\m\n} I$~ for $~\psi\du\m I$~
inspired by the success for purely bosonic non-Abelian
tensors \cite{NRnatensor}, which in turn was inspired by the generalized dimensional
reduction by Scherk-Schwarz \cite{Scherk}.
The field content of our system
is $~(\psi\du{\m\a} I , \chi\du\a{I J} ; ~A\du\m I)$, resembling
the purely bosonic non-Abelian tensor case
$~(B\du{\m\n} I , C\du\m{I J} ,
K^{I J K} ; ~A\du\m I)$.  Similarly to the purely bosonic non-Abelian tensors \cite{NRnatensor}, our field strengths $~\calR\du{\m\n} I$~ for $~\psi\du\m I$~ contains peculiar generalized Chern-Simons terms.

These successful results indicate that we are on the right track for
the formulation of consistent interactions for spin $~3/2$~ fields.  We
have given the first non-trivial interacting model for the nilpotent spinor
charges satisfying $~\{ Q\du\a I, Q\du\b J \} = 0$~ for any arbitrary
space-time dimensions $~D$~ for arbitrary non-Abelian gauge group $~G$.


\begin{theacknowledgments}
  This work is supported in part by NSF Grant \# 0308246.
\end{theacknowledgments}



\bibliographystyle{aipproc}   

\bibliography{sample}

\IfFileExists{\jobname.bbl}{}
 {\typeout{}
  \typeout{******************************************}
  \typeout{** Please run "bibtex \jobname" to optain}
  \typeout{** the bibliography and then re-run LaTeX}
  \typeout{** twice to fix the references!}
  \typeout{******************************************}
  \typeout{}
 }



\end{document}